# Extended Linear Response for Bioanalytical Applications Using Multiple Enzymes

Vladimir Privman,[a]  Oleksandr Zavalov,[a]  Aleksandr Simonian[b]

[a]Department of Physics, Clarkson University, Potsdam, NY 13699, USA
[b]Materials Research and Education Center, Auburn University, Auburn, AL 36849, USA

**Abstract**

We develop a framework for optimizing a novel approach to extending the linear range of bioanalytical systems and biosensors by utilizing two enzymes with different kinetic responses to the input chemical as their substrate. Data for the flow-injection amperometric system devised for detection of lysine based on the function of L-Lysine-alpha-Oxidase and Lysine-2-monooxygenase are analyzed. Lysine is a homotropic substrate for the latter enzyme. We elucidate the mechanism for extending the linear response range and develop optimization techniques for future applications of such systems.

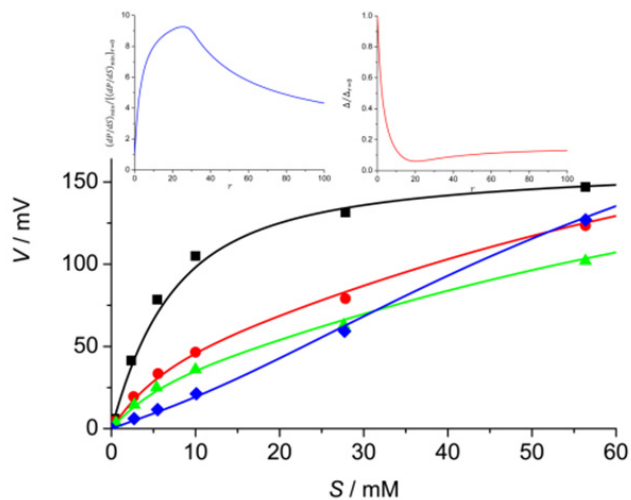

**Introduction**

Recently there has been a renewed interest in using enzymatic reactions and their cascades with pre-designed signal processing features such as, for instance, the response curve for conversion of one or more of the chemical inputs into the reaction output. These developments have been motivated by interest in information and signal processing of the "digital" nature, by utilizing chemical[1-5] and biochemical[6-10] information processing, including that based on enzymatic processes.[11,12] Many of the studied designs have aimed at mimicking binary logic gate functions,[13-15] arithmetic operations,[16] and generally Boolean logic circuits.[14,17,18] New applications have been contemplated for multiple-input analytical systems with response/actuation of the threshold nature.[19,20]

Here we initiate a theoretical development of similar ideas for improving the "proportional" (linear response) characteristics, of interest in many bioanalytical applications including sensors. Indeed, biomolecules, specifically, enzymes, have activity typically varying from batch to batch and over the lifetime of the system exploitation in applications, such as biosensing. Linear response has been a sought after property[21-30] because it offers a convenient way to recalibrate the system by a single or few measurements. We consider an approach to achieve extended linear response by utilizing a combined function of more than a single enzyme. Our present study is based on the analysis of the available data.[31] It was suggested[31] that an enzymatic process with a good linear sensitivity at lower substrate concentrations, and another added allosteric enzymatic process which has an approximately linear response to the same substrate for a range of larger concentrations, can be combined to extend the approximately linear response to cover both ranges.

The kinetic modeling of the type developed here, addresses the actual enzymatic kinetics, which is analyzed by considering the key biocatalytic process steps in detail sufficient to quantify the system response. This offers a novel alternative to possible more phenomenological approaches mentioned later, in the concluding discussion, with the advantage that here we work directly with the parameters that can be controlled to adjust the system behavior, such as the enzyme concentrations. Several studies of bi-enzymatic systems have been reported over the years,[23-30] and therefore initiation of a kinetic modeling approach to their performance design and optimization is timely and will hopefully lead to further research in the future.

We focus on a particular system[31-33] for detection of lysine measured amperometrically by oxygen consumption in a flow-through analyzer. The basic enzymatic reaction is catalyzed by L-Lysine-alpha-Oxidase (LO):

$$\text{L-lysine} + O_2 \xrightarrow{LO} \alpha\text{-keto-}\varepsilon\text{-aminocaproate} + H_2O_2 + NH_3. \qquad (1)$$

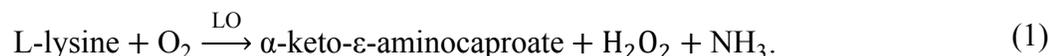

The output of the process is linear in the input concentration of lysine for small enough inputs,[31] saturating for larger concentrations. The second, added process was catalyzed by an allosteric enzyme, Lysine-2-monooxygenase (LMO):

$$\text{L-lysine} + O_2 \xrightarrow{LMO} \delta\text{-aminovaleramide} + CO_2 + H_2O. \qquad (2)$$

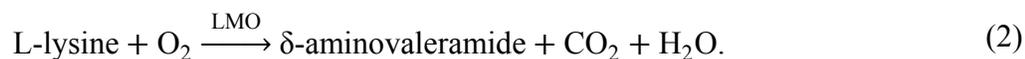

The response curve for the latter process alone is not linear, but rather has a self-promoter sigmoid shape (homotropic response). It was found[31,33] that when both enzymes are present, with LMO added approximately in multiples of the concentration of LO, then the resulting response curve can be made nearly linear over a significantly extended interval of inputs, without substantially reducing the output signal range.

Our aim here is to generally quantify the functioning of such systems and propose a new kinetic approach to improve their sensitivity and degree of linearity. Thus, our model focuses on signal processing rather than on the detailed kinetics of the specific enzymatic processes considered, largely because the latter are not that well known and can be treated approximately for our purposes, as explained in the next section. The last section then details the quality measures and system optimization for improving their bioanalytical response.

**Description of the Enzymatic Process Kinetics**

Generally, for modeling system's response for signal or information processing, or sensing designs involving enzyme-catalyzed processes, it is not always possible to write down a detailed and precise kinetic description of the biochemical processes involved. The reasons for this have been that the available data are noisy and frequently limited, whereas mechanisms of action of many enzymes are not fully understood and can follow several pathways, thus involving many adjustable rate parameters. The latter, rate constants are frequently enzyme batch dependent and usually vary substantially with the physical and chemical conditions of the system. Therefore, one utilizes a simplified kinetic description in terms of rate constants for few key processes, seeking an approximate few-parameter representation of the shape of the system's response curve or surface.

For our non-allosteric enzyme, LO, we use the standard Michaelis-Menten (MM) description:

$$E_1 + S \underset{k_{-1}}{\overset{k_1}{\rightleftarrows}} C_1, \tag{3}$$

$$C_1 \overset{k_2}{\to} E_1 + P. \tag{4}$$

Here the enzyme, the concentration of which is denoted $E_1$, interacts with the substrate (L-lysine) of concentration $S$, to form a complex, $C_1$, which later yields the product, $P$, while restoring the enzyme. The first reaction is usually reversible, but in sensing and information processing applications the reactions are typically driven by the inputs, to have the forward reaction dominate in order to achieve a larger output signal range. Therefore, such systems are frequently in the regime $k_{-1} \approx 0$. Thus, we have a two-variable ($k_{1,2}$) parameterization for an approximate description of the LO catalyzed kinetics.

Note that all the chemical concentrations here, $E_1(t)$, etc., are time dependent. They are also spatially varying along the flow. However, the enzymatic processes (for both LO and also LMO) are in their steady states, and therefore we can use a representative product, $P(t)$, concentration calculated as a function of time as a certain "measurement time," $t = t_\mathrm{m}$, typical for the experiment, which here was $t_\mathrm{m} = 120$ sec,[32-34] and otherwise ignore spatial variation along the flow. This is justified by the fact that our model is anyway simplified, aimed at "response shape fitting" in the sense explained earlier, and the actual signal measured, to be denoted $V$, is not one of the concentrations of the product chemicals but rather the rate of the oxygen consumption, measured amperometrically. The conversion of the "product" shown in our relations to the actually measured signal is not known and can depend on the chemical conditions, specifically, the pH,[32] on the particular setup used, and on the enzyme batch, etc. Therefore, we simply regard the conversion factor as another adjustable parameter:

$$V = \gamma P(t_\mathrm{m}). \quad (5)$$

With these conventions, we can set up rate equations, given in the Supporting Information (appended at the end of this preprint), with the LMO kinetics added as described shortly, and thus model the available data. This is illustrated in Figure 1 (all the figures are at the collated at the end of the main text, see Pages 13-17), where the data fit result is shown for an LO only experiment, with the initial values $E_1(0) = 0.26$ µM, $S(0)$ varying from 0 up to about 55 mM, and the fitted constants $k_1 = 2.1 \times 10^{-3}$ mM$^{-1}$sec$^{-1}$, $k_2 = 1.0 \times 10^{-5}$ sec$^{-1}$, $\gamma = 145$ mV/mM.

For LMO, of concentration denoted $E_2(t)$, we have to consider the fact that L-lysine seems to have a self-promoter property, as can be seen from the data in Figure 1. Presumably this indicates some sort of allostericity. However, the detailed mechanism has not been reported in the literature. Therefore, we use the most straightforward modeling approach to conventional self-promoter allostericity, i.e., a variant of the kinetics[35] of the type

$$E_2 + S \underset{k_{-3}}{\overset{k_3}{\rightleftarrows}} E_2^a, \quad (6)$$

$$E_2^a + S \underset{k_{-4}}{\overset{k_4}{\rightleftarrows}} C_2, \quad (7)$$

– 5 –

$$C_2 \xrightarrow{k_5} E_2^a + P. \tag{8}$$

This is of course a significantly simplified description whereby we assume that some of the substrate is used up to produce a more active form of LMO: $E_2^a$ (we lump all such possible forms into a single concentration to minimize the number of parameters). We then assume that this effective "active" enzyme functions according to the MM scheme similar to Equations (3-4), with similar assumptions, for example, here we set $k_{-4} = 0$ (and also $k_{-3} = 0$). This approach ignores several possible kinetic pathways and, as mentioned, lumps some processes into effective single steps. However, it allows us to limit the fitting to only three new parameters. (We assume that the product-to-signal conversion factor, $\gamma$, is the same and thus avoid using it as another adjustable parameter here.) In fact, the products of the LMO reaction are not the same as for LO, see Equations (1)-(2), but we use the same notation, $P$, because both processes affect the oxygen consumption rate which is the actual measured quantity.

An illustration of a data fit for an LMO-only system is given in Figure 1. Here we used $E_2(0) = 0.25$ μM, $S(0)$ varying in the same range as before, and the fitted constants $k_3 = 0.39 \times 10^{-3}$ mM$^{-1}$sec$^{-1}$, $k_4 = 0.47 \times 10^{-3}$ mM$^{-1}$sec$^{-1}$, $k_5 = 1.2 \times 10^{-3}$ sec$^{-1}$. The success of the model fit here and, as described shortly, for situations with more than one enzyme (with the same fitting parameter values) provides a new evidence that LMO is indeed allosteric, with L-lysine a homotropic substrate.

Data vs. model with two enzymes present, are given in Figure 2. Good degree of consistency was found for all the initial LMO:LO ratios for the reported[31,36] data. Note that the functioning of the two enzymes is coupled via the L-lysine substrate concentration. This can be seen in the set of the rate equation given in the Supporting Information, see Equations (S1)-(S5), which allow the calculation of $V(t)$ which in turn gives the measured signal at time $t = t_m$. These rate equations were used to fit the LO only and LMO only data (see Figure 1), as well as offered a reasonable model description of the data with multiples of LMO added to the LO system (see Figure 2). The initial concentrations of the enzymes for the experiments shown in Figure 2 were taken as $E_1(0) = 0.29$ μM for LO, and $E_2(0) = r \times 0.26$ μM for LMO, with $r = 1, 8, 12, 29$. They are not precisely equal for $r = 1$ in terms of molar concentrations, but they were the same in terms of the amounts of the enzymes per unit mass of the gel on which they were immobilized.[31,33]

**Evaluation and Optimization of the Extended Linear Response**

In order to analyze the system response with both enzymes present, let us first comment on the observation that the two enzymes obviously compete for the substrate and thus affect each other's activity, as can be seen from Figures 1 and 2. Here we focus on the experimentally realized situation with LMO present initially as (approximately) a multiple, $r$, of the LO concentration (other combinations can be analyzed similarly). The value of $r$ was increased[31] up to 29. Mere homotropicity of the substrate with respect to the added enzyme might not always be sufficient to produce the desired extended, semi-linear response by the two-enzyme competition. The bi-enzyme system must be properly tuned by the enzyme selection, experimental studies, modeling, and parameter selection, to function in this regime.

For our particular system an obvious quantity to consider for optimization is the selection of the value of $r$, assuming that it is the only parameter being easily adjusted, though our discussion below is generally applicable to other systems, situations, and parameter variation options. We first consider the differential sensitivity of the system over the desired input range, assumed to be $S(0) = 0$ to 60 mM. The LO-only output signal, proportional to $P = P(t_m)$, is approximately linear for inputs, $S = S(0)$, but only in the range between 0 and 5 mM. This is seen in Figure 1, whereas in Figure 2 we observe that the response of the modified system with $r = 29$ is approximately linear for our target input range, 0 to 60 mM. Obviously, there is a tradeoff that the overall sensor sensitivity drops due to the modification that resulted in a larger approximately linear response range. In Figure 3, we plot the differential sensitivity,

$$\frac{dP}{dS} = \frac{dP(t_m)}{dS(0)}, \tag{9}$$

over the target input range. The goal of the optimization could be to have this function's *minimum value* as large as possible, and inspection of Figure 3 shows that $r = 29$ could indeed be a good choice.

A more careful analysis of the differential sensitivity is shown in Figure 4, where the minimal value of the derivative in Equation (9), $(dP/dS)_{\min}$, over the target input signal range is plotted normalized per its value at $r = 0$ in order to have dimensionless units. Figure 4 suggests that, based on the presently considered criterion alone, the values of $r$ from approximately 16 to 32 represent an optimal range for this parameter selection.

However, in addition to improving sensitivity, other criteria for sensor functioning are of importance. Specifically, the present data were considered[31] with the aim at improving the linearity of the response. To analyze linearity, we consider the root-mean-square deviation of the slope of the response curve from the average slope, as detailed in the Supporting Information. The measure, $\Delta$, of the response curve linearity is defined, see Equations (S6)-(S8), as the root-mean-square deviation normalized per the average slope over the input range from 0 to $S_{\max} = 60$ mM. For best linearity of the response curve, we have to minimize this measure, which is plotted as a function of $r$ in Figure 5 (normalized per its value at $r = 0$). Inspection of Figure 5 suggests that values from approximately 14 to 32 represent an optimal range for the parameter $r$ selection based on the "linearity" criterion. Thus, both criteria in the present system yield very similar ranges for $r$. Interestingly, the actual minimum of the linearity measure (Figure 5) is close to $r = 20$, consistent with the estimate quoted in the experimental work[31] based on the data analyzed here as well as other data from different experiments. On the other hand, the maximum of the differential-sensitivity criterion measure (Figure 4) is close to $r = 25$. A larger degree of tradeoff may be expected in other situations if additional optimization measures are considered.

**Conclusion**

We reported an approach which allows optimization of the linear regime of bioanalytical systems based on functioning of two enzymes with different kinetic mechanisms, in the regime when they are competing for the same substrate. An advantage of the present modeling strategy as compared to other more phenomenological approaches,[37] such as the use of Hill functions,[38] or principal component analysis[39,40] seeking correlations between various parameters of complex systems and processes, has been in that our kinetic description allows to identify the effect of varying those system parameters which are not dependent on the inputs and can be relatively easily controlled. Here we focused on the concentration of one of the enzymes which was earlier found to have a profound effect on the response curve.

We considered two measures of the response quality: differential sensitivity, the minimum value of which should be made as large as possible over the target range of inputs, and a measure of linearity, which should be minimized. In the particular example considered both criteria can be consistently satisfied. It is hoped that the developed ideas will stimulate additional research on bi-enzyme[23-30,39,41-43] and other systems[44-46] with similar response characteristics.

**Supporting Information**

Additional information as noted in text. This material is available free of charge via the Internet at http://pubs.acs.org (as well as appended at the end of this preprint).

**Acknowledgements**

We wish to thank Prof. E. Katz and Dr. J. Halámek for helpful input and discussions. The team at Clarkson University acknowledges support by the National Science Foundation (grants CBET-1066397 and CCF-1015983). The team at Auburn University acknowledges support from the USDA-CSREES (grant 2006-34394-16953).

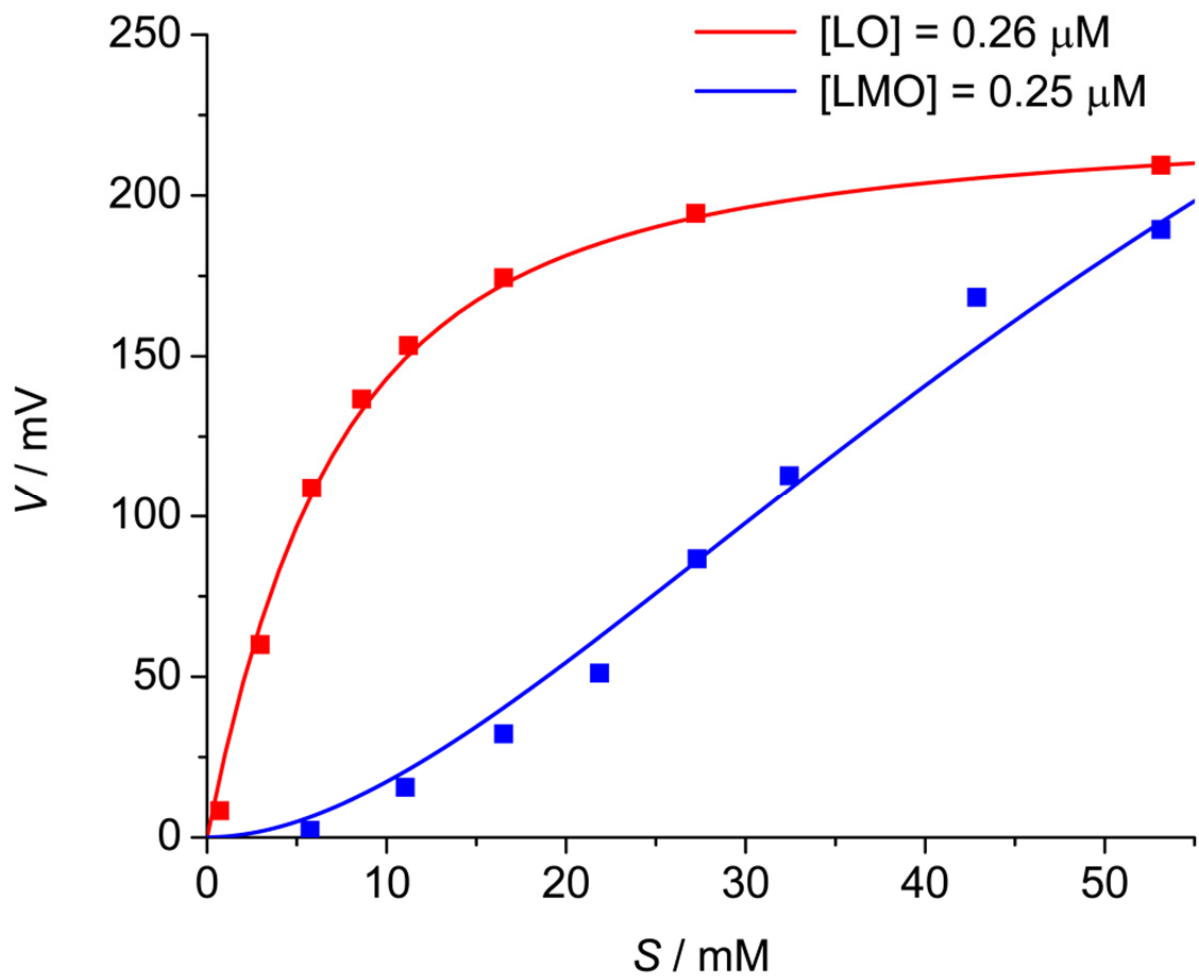

**Figure 1:** Illustration of data fitting for the LO only (red curve and symbols) and LMO only (coded in blue) systems, with the model parameters as described in the text.

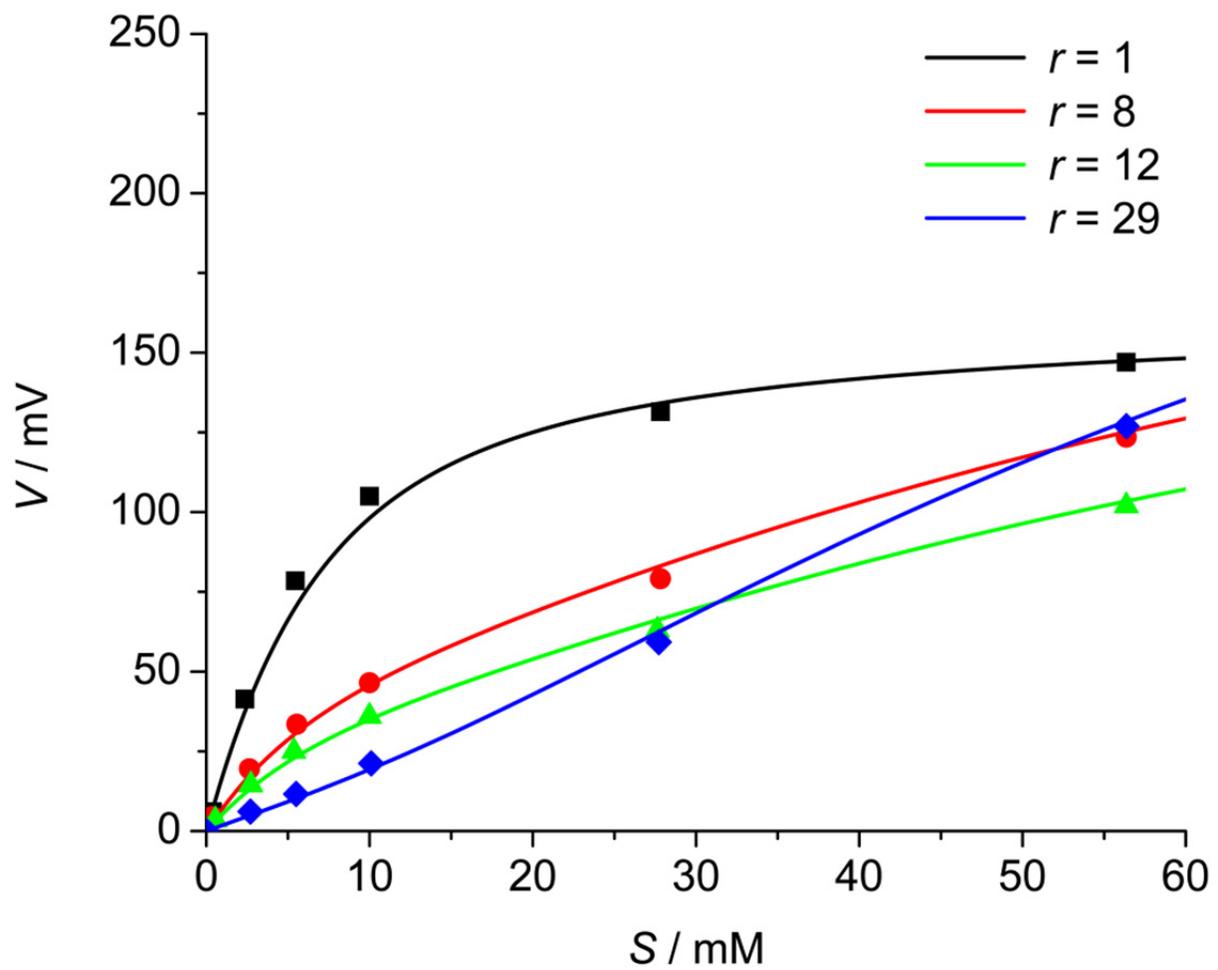

**Figure 2:** Illustration of data fitting for LMO:LO initially present at ratios $r = 1, 8, 12, 29$, as described in the text.

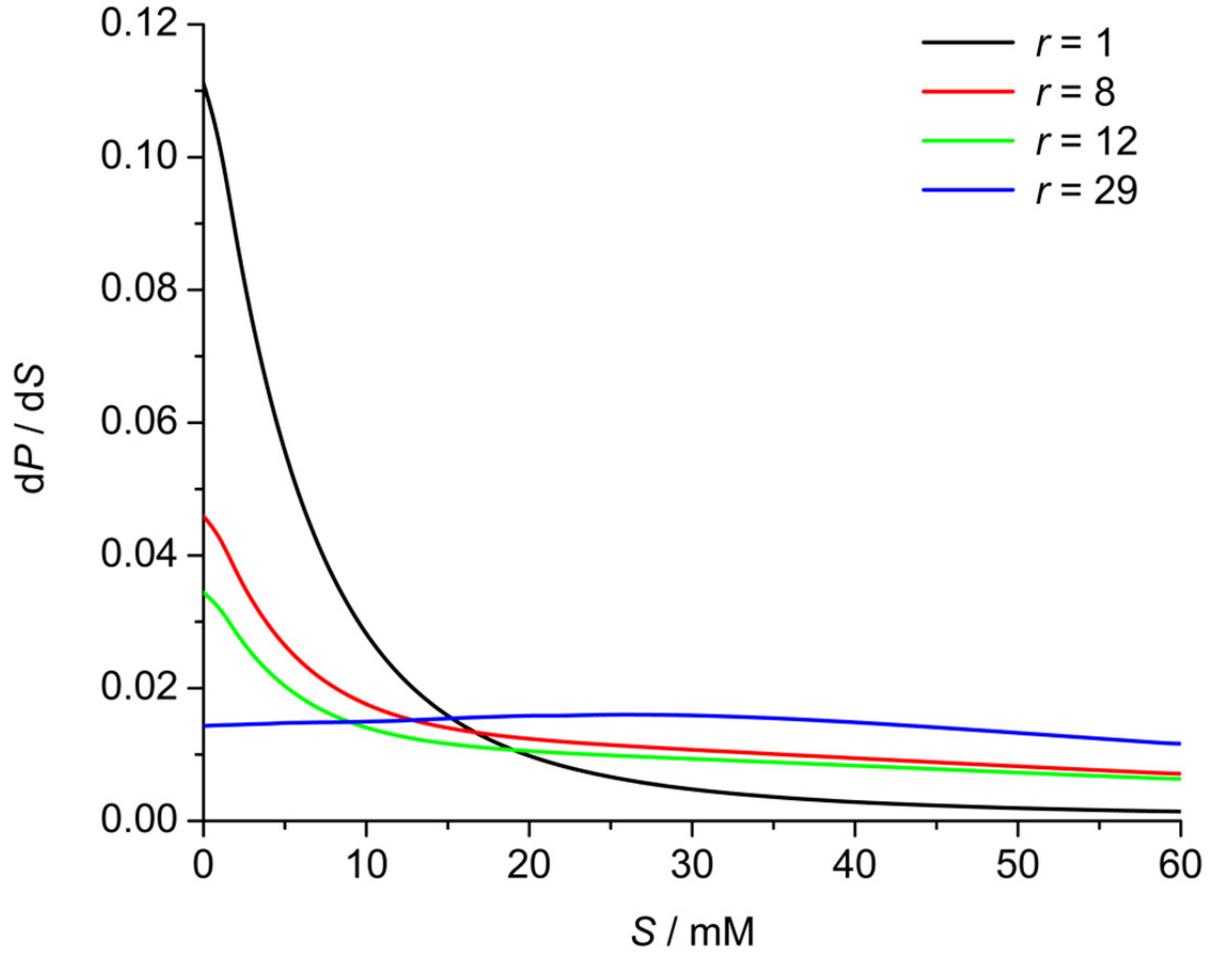

**Figure 3:** Differential sensitivity calculated over the desired input range, 0 to 60 mM, for various values of the ratio, $r$.

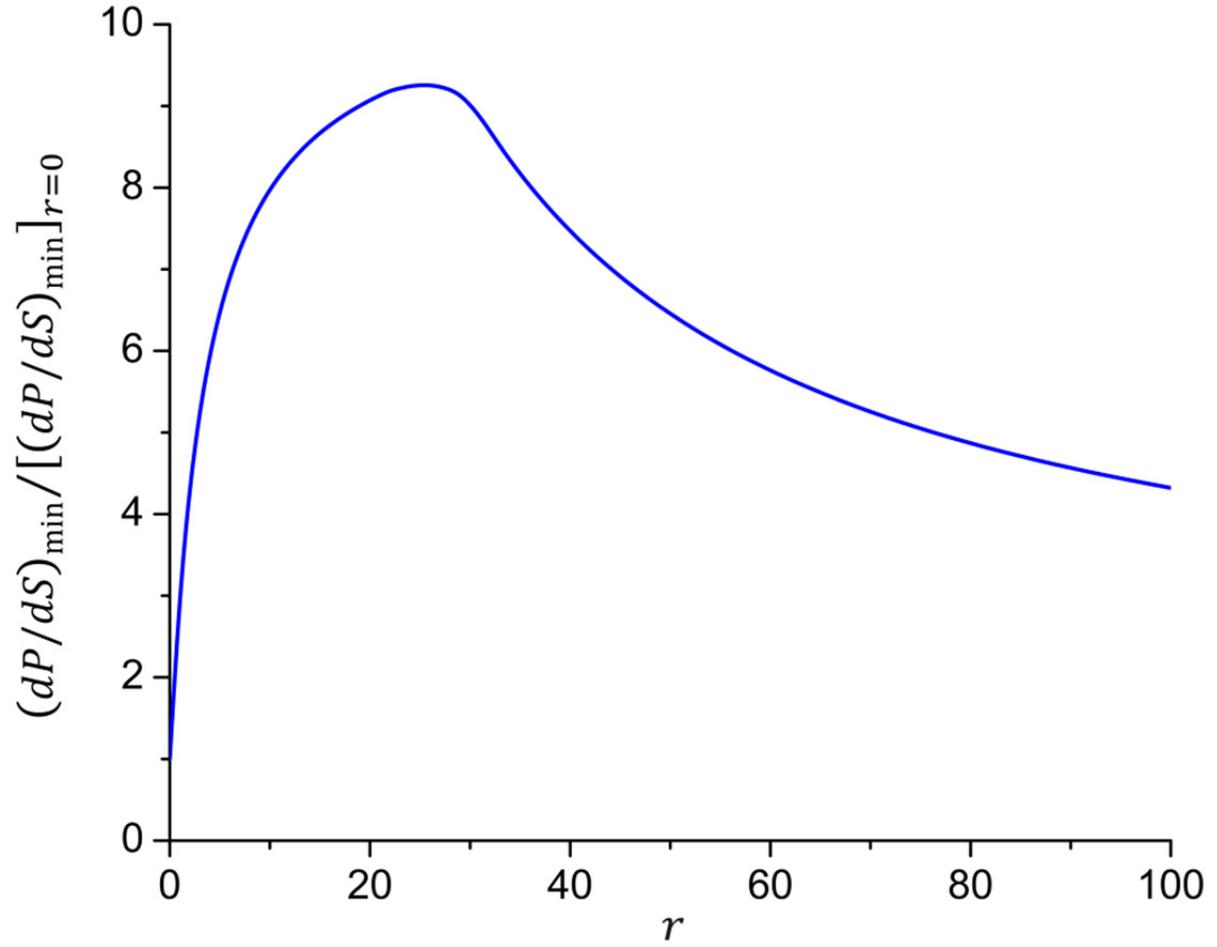

**Figure 4:** The minimum value of the differential sensitivity over the desired input range, 0 to 60 mM, normalized per its value at $r = 0$, plotted as a function of $r$.

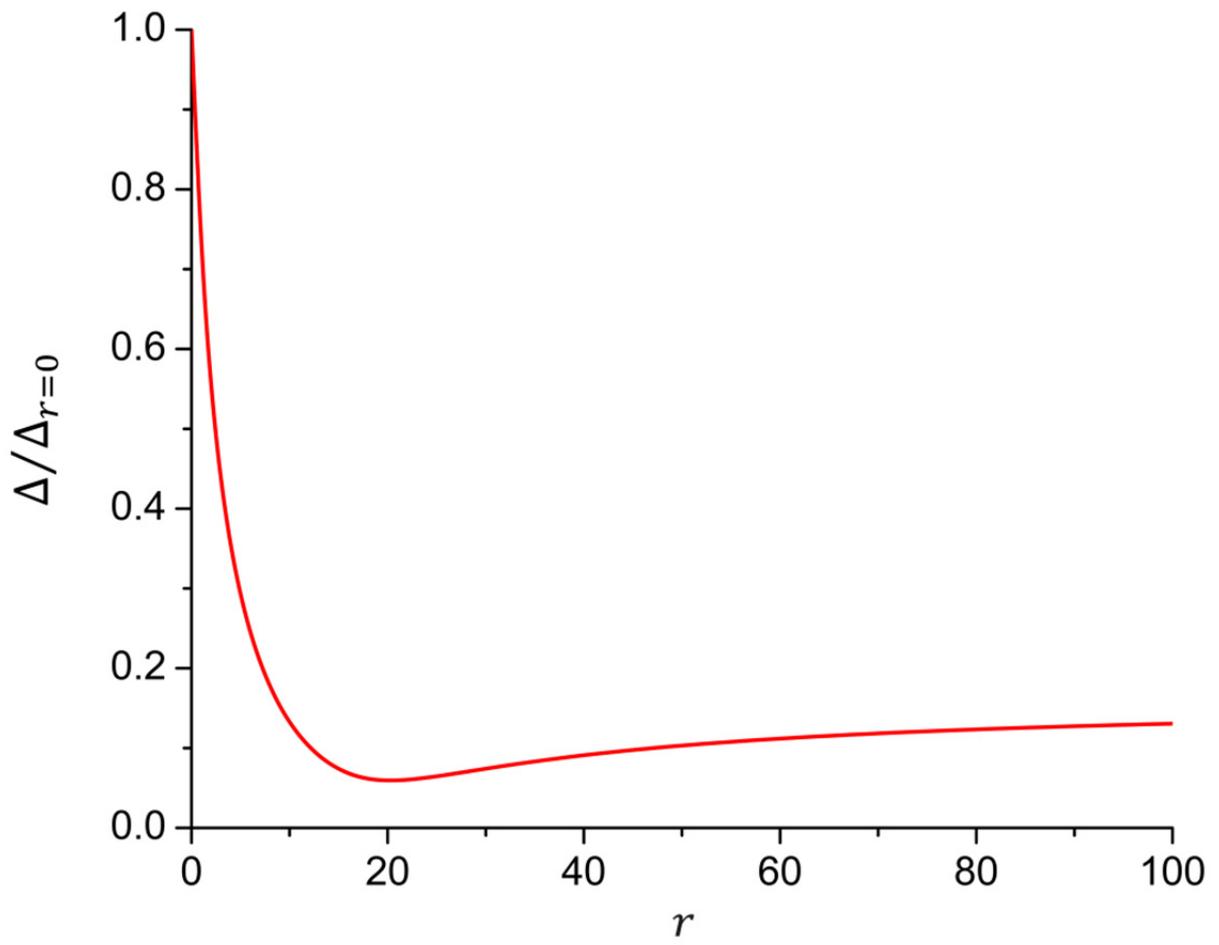

**Figure 5:** Measure of the response curve linearity.

# Extended Linear Response for Bioanalytical Applications Using Multiple Enzymes

Vladimir Privman,[a] Oleksandr Zavalov,[a] Aleksandr Simonian[b*]

[a]Department of Physics, Clarkson University, Potsdam, NY 13699, USA
[b]Materials Research and Education Center, Auburn University, Auburn, AL 36849, USA

*Corresponding author: phone (334) 844-4485, e-mail als@auburn.edu

## SUPPORTING INFORMATION

**Rate Equations**

The set of the rate equations which correspond to Equations (3)-(8) is

$$\frac{dE_1(t)}{dt} = -k_1 SE_1 + k_2 C_1, \tag{S1}$$

$$\frac{dS(t)}{dt} = -k_1 SE_1 - k_3 SE_2 - k_4 SE_2^a, \tag{S2}$$

$$\frac{dE_2(t)}{dt} = -k_3 SE_2, \tag{S3}$$

$$\frac{dE_2^a(t)}{dt} = k_3 SE_2 - k_4 SE_2^a + k_5 C_2, \tag{S4}$$

$$\frac{dV(t)}{dt} = \gamma(k_2 C_1 + k_5 C_2), \tag{S5}$$

with all the concentrations on the right-hand sides time-dependent, and with $C_1(t) = E_1(0) - E_1(t)$ and $C_2(t) = E_2(0) - E_2(t) - E_2^a(t)$.

**Measure of the Linearity of the Response**

The root-mean-square deviation of the slope of the response curve from the average slope is proportional to the following quantity,

$$\sqrt{\int_0^{S_{max}} \left[\frac{dP(S)}{dS} - \frac{P(S_{max})-P(0)}{S_{max}}\right]^2 dS}, \quad (S6)$$

where the average slope over the input range from 0 to $S_{max}$ is

$$\frac{|P(S_{max})-P(0)|}{S_{max}}. \quad (S7)$$

The measure $\Delta$ can be defined as the room-mean-squared deviation normalized per the average slope,

$$\Delta = \frac{\sqrt{\int_0^{S_{max}} \left[\frac{dP(S)}{dS} - \frac{P(S_{max})-P(0)}{S_{max}}\right]^2 dS}}{\frac{|P(S_{max})-P(0)|}{S_{max}}}. \quad (S8)$$